\documentclass[conference, final]{IEEEtran}
\usepackage[nolist]{acronym}
\usepackage{tikz}
\usetikzlibrary{decorations.pathreplacing,calligraphy}
\usepackage{pgfplots}
\usepgfplotslibrary{groupplots}
\usepackage{graphicx}
\usepackage{subfloat}
\usepackage[font=footnotesize]{caption}
\usepackage{subcaption}
\usepackage{float}
\usepackage{array}
\pgfplotsset{compat=newest}
\usepackage{amsmath, amsbsy, amssymb}
\usepackage{tikzscale}
\usepackage{color}
\usepackage{epigraph}
\usepackage{circuitikz}
\usepackage{multirow}
\usepackage{booktabs}
\usepackage[ruled]{algorithm2e}
\usepackage{adjustbox}
\usepackage[group-separator={,}]{siunitx}
\definecolor{darkgreen}{rgb}{0.125,0.5,0.169}
\usetikzlibrary{shapes,arrows,fadings,chains}
\usepackage{marvosym}
\usepackage{bbm}
\usepackage{mathtools}
\usetikzlibrary{positioning,fit,calc}
\usetikzlibrary{calc,patterns,angles,quotes}
\usetikzlibrary{shapes.geometric, arrows}
\usetikzlibrary{fit}
\usetikzlibrary{shapes.multipart}
\usetikzlibrary{positioning,fit,calc}
\usetikzlibrary{calc, positioning}
\usetikzlibrary{decorations}
\usetikzlibrary{decorations.pathreplacing}
\usetikzlibrary{arrows.meta,shapes.arrows}
\usetikzlibrary{shapes}
\usetikzlibrary{positioning,fit,backgrounds}
\usepgfplotslibrary{groupplots}
\pgfplotsset{compat=1.13}
\usetikzlibrary{external}
\tikzset{>=latex}
\usepackage{tikzit}

\begin{acronym}
 \acro{CSI}{channel state information}
 \acro{UE}{user equipment}
 \acro{UL}{uplink}
 \acro{BS}{basestation}
 \acro{TDD}{time division duplex}
 \acro{FDD}{frequency division duplex}
 \acro{ECC}{error-correcting code}
 \acro{MLD}{maximum likelihood decoding}
 \acro{HDD}{hard decision decoding}
 \acro{IF}{intermediate frequency}
 \acro{RF}{radio frequency}
 \acro{SDD}{soft decision decoding}
 \acro{NND}{neural network decoding}
 \acro{CNN}{convolutional neural network}
 \acro{ML}{maximum likelihood}
 \acro{GPU}{graphical processing unit}
 \acro{BP}{belief propagation}
 \acro{LTE}{Long Term Evolution}
 \acro{BER}{bit error rate}
 \acro{DER}{detection error rate}
 \acro{SNR}{signal-to-noise-ratio}
 \acro{ReLU}{rectified linear unit}
 \acro{BPSK}{binary phase shift keying}
 \acro{QPSK}{quadrature phase shift keying}
 \acro{AWGN}{additive white Gaussian noise}
 \acro{MSE}{mean squared error}
 \acro{LLR}{log-likelihood ratio}
 \acro{MAP}{maximum a posteriori}
 \acro{NVE}{normalized validation error}
 \acro{BCE}{binary cross-entropy}
 \acro{CE}{cross-entropy}
 \acro{BLER}{block error rate}
 \acro{SQR}{signal-to-quantisation-noise-ratio}
 \acro{MIMO}{multiple-input multiple-output}
 \acro{OFDM}{orthogonal frequency division multiplex}
 \acro{RF}{radio frequency}
 \acro{LOS}{line of sight}
 \acro{NLoS}{non-line of sight}
 \acro{NMSE}{normalized mean squared error}
 \acro{CFO}{carrier frequency offset}
 \acro{SFO}{sampling frequency offset}
 \acro{IPS}{indoor positioning system}
 \acro{TRIPS}{time-reversal IPS}
 \acro{RSSI}{received signal strength indicator}
 \acro{MIMO}{multiple-input multiple-output}
 \acro{ENoB}{effective number of bits}
 \acro{AGC}{automated gain control}
 \acro{ADC}{analog to digital converter}
 \acro{ADCs}{analog to digital converters}
 \acro{FB}{front bandpass}
 \acro{FPGA}{field programmable gate array}
 \acro{JSDM}{Joint Spatial Division and Multiplexing}
 \acro{NN}{neural network}
 \acro{IF}{intermediate frequency}
 \acro{LoS}{line-of-sight}
 \acro{NLoS}{non-line-of-sight}
 \acro{DSP}{digital signal processing}
 \acro{AFE}{analog front end}
 \acro{SQNR}{signal-to-quantisation-noise-ratio}
 \acro{SINR}{signal-to-interference-noise-ratio}
 \acro{ENoB}{effective number of bits}
 \acro{AGC}{automated gain control}
 \acro{PCB}{printed circuit board}
 \acro{EVM}{error vector mangnitude}
 \acro{CDF}{cumulative distribution function}
 \acro{MRC}{maximum ratio combining}
 \acro{MRP}{maximum ratio precoding}
 \acro{MRT}{maximum ratio transmission}
 \acro{DeepL}{deep-learning}
 \acro{DL}{deep learning}
 \acro{SISO}{single-input single-output}
 \acro{SGD}{stochastic gradient descent}
 \acro{CP}{cyclic prefix}
 \acro{MISO}{Multiple Input Single Output}
 \acro{LMMSE}{linear minimum mean square error}
 \acro{ZF}{zero forcing}
 \acro{USRP}{universal software radio peripheral}
 \acro{RNN}{recurrent neural network}
 \acro{GRU}{gated recurrent unit}
 \acro{LSTM}{long short-term memory}
 \acro{NTM}{neural turing machine}
 \acro{DNC}{differentiable neural computer}
 \acro{TCN}{temporal convolutional network}
 \acro{FCL}{fully connected layer}
 \acro{MANN}{memory augmented neural network}
 \acro{FNN}{feedforward neural network}
 \acro{DNN}{dense neural network}
 \acro{FIR}{finite impulse response}
 \acro{BPTT}{back-propagation through time}
 \acro{GAN}{generative adversarial network}
 \acro{ELU}{exponential linear unit}
 \acro{tanh}{hyperbolic tangent}
 \acro{BICM}{bit-interleaved coded modulation}
 \acro{OTA}{over-the-air}
 \acro{IM}{intensity modulation}
 \acro{DD}{direct detection}
 \acro{RL}{reinforcement learning}
 \acro{SDR}{software-defined radio}
 \acro{WGAN}{Wasserstein generative adversarial network}
 \acro{BMD}{bit-metric decoding}
 \acro{BMI}{bit-wise mutual information}
 \acro{LDPC}{low-density parity-check}
 \acro{IDD}{iterative demapping and decoding}
 \acro{IEDD}{iterative equalization, demapping and decoding}
 \acro{JSD}{Jensen-Shannon divergence}
 \acro{MMSE}{minimum mean square error}
 \acro{FFT}{fast Fourier transform}
 \acro{IFFT}{inverse fast Fourier transform}
 \acro{QAM}{quadrature amplitude modulation}
 \acro{EMD}{earth mover's distance}
 \acro{TDL}{tapped delay line}
 \acro{KL}{Kullback-Leibler}
 \acro{PRACH}{physical random access channel}
 \acro{URLLC}{ultra-reliable low-latency communication}
 \acro{ANOMA}{asynchronous non-orthogonal multiple access}
 \acro{FEC}{forward error correction}
 \acro{NOMA}{non-orthogonal medium access}
 \acro{MTC}{machine-type communications}
 \acro{mMTC}{massive machine-type communications}
 \acro{MCS}{modulation and coding scheme}
 \acro{PAPR}{peak-to-average power ratio}
 \acro{MAC}{medium access control}
 \acro{STO}{sampling time offset}
 \acro{STE}{straight-through estimator}
 \acro{PHY}{physical}
 \acro{CCE}{categorical cross-entropy}
 \acro{IoT}{Internet of Things}
 \acro{CCDF}{complementary cumulative distribution function}
 \acro{CRC}{cyclic redundancy check}
 \acro{ACLR}{adjacent channel leakage ratio}
\acro{MD}{missed detection}
\acro{FA}{false alarm}
\acro{FAR}{false alarm rate}
\acro{BCJR}{Bahl-Cocke-Jelinek-Raviv}
\acro{BA}{beam alignment}
\acro{DRL}{deep reinforcement learning}
\acro{MDP}{markov decision process}
\acro{POMDP}{partially observable Markov decision process}
\acro{ULA}{uniform linar array}
\acro{UPA}{uniform planar array}
\acro{PPO}{proximal policy optimization}
\acro{AoA}{angle-of-arrival}
\acro{AoD}{angle-of-departure}
\acro{IA}{initial access}
\acro{OMP}{orthogonal matching pursuit}
\acro{CU}{control unit}
\acro{SA}{standalone}
\acro{NSA}{non-standalone}
\acro{BF}{beamforming}
\acro{BT}{beam training}
\acro{RA}{random access}
\acro{RS}{reference signal}
\acro{SSB}{synchronization signal block}
\acro{TX}{transmitter}
\acro{RX}{receiver}
\end{acronym}

\definecolor{mittelblau}{RGB}{0, 126, 198}
\definecolor{violettblau}{cmyk}{0.9, 0.6, 0, 0}
\definecolor{rot}{RGB}{238, 28 35}
\definecolor{apfelgruen}{RGB}{140, 198, 62}
\definecolor{gelb}{RGB}{1, 221, 0}
\definecolor{orange}{RGB}{244, 111, 33}
\definecolor{pink}{RGB}{237, 0, 140}
\definecolor{lila}{RGB}{128, 10, 145}
\definecolor{hellgrau}{RGB}{224, 224, 224}
\definecolor{mittelgrau}{RGB}{128, 128, 128}
\definecolor{dunkelgrau}{RGB}{80,80,80}
\definecolor{anthrazit}{RGB}{19, 31, 31}

\pgfplotscreateplotcyclelist{corporate colours markers}{%
rot, every mark/.append style={fill=.!80!rot},mark=*\\%
mittelblau, every mark/.append style={fill=.!80!mittelblau},mark=square*\\%
apfelgruen, every mark/.append style={fill=.!80!apfelgruen},mark=triangle*\\%
orange, mark=star\\%
pink, every mark/.append style={fill=.!80!pink},mark=diamond*\\%
violettblau, every mark/.append style={fill=.!80!violettblau},mark=otimes*\\%
lila, mark=|\\%
gelb, every mark/.append style={fill=.!80!gelb},mark=pentagon*\\%
hellgrau, mark=text,text mark=p\\%
anthrazit, mark=text,text mark=a\\%
}

\pgfplotscreateplotcyclelist{corporate colours markers double}{%
rot, every mark/.append style={fill=.!80!rot},mark=*\\%
rot, dashed, every mark/.append style={fill=.!80!rot,solid},mark=*\\%
mittelblau, every mark/.append style={fill=.!80!mittelblau},mark=square*\\%
mittelblau, dashed, every mark/.append style={fill=.!80!mittelblau,solid},mark=square*\\%
apfelgruen, every mark/.append style={fill=.!80!apfelgruen},mark=triangle*\\%
apfelgruen, dashed, every mark/.append style={fill=.!80!apfelgruen,solid},mark=triangle*\\%
orange, mark=star\\%
orange, dashed, mark=star\\%
pink, every mark/.append style={fill=.!80!pink},mark=diamond*\\%
pink, dashed, every mark/.append style={fill=.!80!pink,solid},mark=diamond*\\%
violettblau, every mark/.append style={fill=.!80!violettblau},mark=otimes*\\%
violettblau, dashed, every mark/.append style={fill=.!80!violettblau,solid},mark=otimes*\\%
lila, mark=|\\%
lila, dashed, mark=|\\%
gelb, every mark/.append style={fill=.!80!gelb},mark=pentagon*\\%
gelb, dashed, every mark/.append style={fill=.!80!gelb,solid},mark=pentagon*\\%
}

\pgfplotsset{
  colormap/magma/.style={%
    /pgfplots/colormap={magma}{%
      rgb=(0.001462, 0.000466, 0.013866)
      rgb=(0.035520, 0.028397, 0.125209)
      rgb=(0.102815, 0.063010, 0.257854)
      rgb=(0.191460, 0.064818, 0.396152)
      rgb=(0.291366, 0.064553, 0.475462)
      rgb=(0.384299, 0.097855, 0.501002)
      rgb=(0.475780, 0.134577, 0.507921)
      rgb=(0.569172, 0.167454, 0.504105)
      rgb=(0.664915, 0.198075, 0.488836)
      rgb=(0.761077, 0.231214, 0.460162)
      rgb=(0.852126, 0.276106, 0.418573)
      rgb=(0.925937, 0.346844, 0.374959)
      rgb=(0.969680, 0.446936, 0.360311)
      rgb=(0.989363, 0.557873, 0.391671)
      rgb=(0.996580, 0.668256, 0.456192)
      rgb=(0.996727, 0.776795, 0.541039)
      rgb=(0.992440, 0.884330, 0.640099)
      rgb=(0.987053, 0.991438, 0.749504)
    },
  },
  colormap/inferno/.style={%
    /pgfplots/colormap={inferno}{%
      rgb=(0.001462, 0.000466, 0.013866)
      rgb=(0.037668, 0.025921, 0.132232)
      rgb=(0.116656, 0.047574, 0.272321)
      rgb=(0.217949, 0.036615, 0.383522)
      rgb=(0.316282, 0.053490, 0.425116)
      rgb=(0.410113, 0.087896, 0.433098)
      rgb=(0.503493, 0.121575, 0.423356)
      rgb=(0.596940, 0.154848, 0.398125)
      rgb=(0.688653, 0.192239, 0.357603)
      rgb=(0.775059, 0.239667, 0.303526)
      rgb=(0.851384, 0.302260, 0.239636)
      rgb=(0.912966, 0.381636, 0.169755)
      rgb=(0.956852, 0.475356, 0.094695)
      rgb=(0.981895, 0.579392, 0.026250)
      rgb=(0.987464, 0.690366, 0.079990)
      rgb=(0.973088, 0.805409, 0.216877)
      rgb=(0.947594, 0.917399, 0.410665)
      rgb=(0.988362, 0.998364, 0.644924)
    },
  },
  colormap/plasma/.style={%
    /pgfplots/colormap={plasma}{%
      rgb=(0.050383, 0.029803, 0.527975)
      rgb=(0.186213, 0.018803, 0.587228)
      rgb=(0.287076, 0.010855, 0.627295)
      rgb=(0.381047, 0.001814, 0.653068)
      rgb=(0.471457, 0.005678, 0.659897)
      rgb=(0.557243, 0.047331, 0.643443)
      rgb=(0.636008, 0.112092, 0.605205)
      rgb=(0.706178, 0.178437, 0.553657)
      rgb=(0.768090, 0.244817, 0.498465)
      rgb=(0.823132, 0.311261, 0.444806)
      rgb=(0.872303, 0.378774, 0.393355)
      rgb=(0.915471, 0.448807, 0.342890)
      rgb=(0.951344, 0.522850, 0.292275)
      rgb=(0.977856, 0.602051, 0.241387)
      rgb=(0.992541, 0.687030, 0.192170)
      rgb=(0.992505, 0.777967, 0.152855)
      rgb=(0.974443, 0.874622, 0.144061)
      rgb=(0.940015, 0.975158, 0.131326)
    },
  },
}

\tikzstyle{test 1}=[fill=white, draw=black, shape=circle]
\tikzstyle{plus}=[fill=white, draw=black, shape=circle, scale=0.5]

\tikzstyle{arrow 1}=[->, -{triangle 45[length=3mm]}, draw={rgb,255: red,62; green,48; blue,255}, fill={rgb,255: red,62; green,48; blue,255}, ultra thick]
\tikzstyle{box bs}=[-, draw=mittelblau, fill = mittelblau, fill opacity=0.1, ultra thick]
\tikzstyle{box ue}=[-, draw=orange, fill = orange, fill opacity=0.1, ultra thick]
\tikzstyle{box controller}=[-, fill=rot, draw=rot, fill opacity=0.2, ultra thick, rounded corners]
\tikzstyle{box parameter}=[-, fill=apfelgruen!75, draw=apfelgruen, thick]
\tikzstyle{arrow 2}=[->, -{triangle 45[length=3mm]}, semithick]
\tikzstyle{box3}=[-, fill=white]
\tikzstyle{arrow 3}=[->, -{triangle 45[length=3mm]}, semithick, double]
\tikzstyle{arrow 4}=[->, -{triangle 45[length=3mm]}, thick]
\tikzstyle{arrow 5}=[-, thick]
\tikzstyle{arrow 6}=[->, -{triangle 45[length=3mm]}, thick, double]
\tikzstyle{arrow 7}=[-, thick, double]
\tikzstyle{arrow 8}=[-, semithick, double]
\tikzstyle{arrow 9}=[->, -{triangle 45[length=3mm]}, draw={rot}, fill={rot}, ultra thick]
\tikzstyle{arrow 10}=[-, draw={rot}, fill={rot}, ultra thick]

\tikzstyle{box 1}=[-, draw={mittelblau}, fill={mittelblau}, fill opacity=0.1, ultra thick]
\tikzstyle{box 2}=[-, draw={orange}, fill opacity=0.2, ultra thick]
\tikzstyle{box 3}=[-, fill={rot}, draw=rot, fill opacity=0.2, ultra thick, rounded corners]

\IEEEoverridecommandlockouts

\begin{document}

\title{Deep Learning Based Adaptive Joint mmWave Beam Alignment}

\author{\IEEEauthorblockN{Daniel Tandler\IEEEauthorrefmark{1}, Marc Gauger\IEEEauthorrefmark{1}, Ahmet Serdar Tan\IEEEauthorrefmark{2}, Sebastian D\"orner\IEEEauthorrefmark{1} and Stephan ten Brink\IEEEauthorrefmark{1}}

\IEEEauthorblockA{
 \IEEEauthorrefmark{1}Institute of Telecommunications, University of Stuttgart, Pfaffenwaldring 47, 70659 Stuttgart, Germany \\
\{tandler,gauger,doerner, tenbrink\}@inue.uni-stuttgart.de\\
\IEEEauthorrefmark{2} InterDigital Europe, London, United Kingdom
}

}

\maketitle
\begin{abstract}
The challenging propagation environment, combined with the hardware limitations of mmWave systems, gives rise to the need for accurate initial access beam alignment strategies with low latency and high achievable beamforming gain.
Much of the recent work in this area either focuses on one-sided beam alignment, or, joint beam alignment methods where both sides of the link perform a sequence of fixed channel probing steps. Codebook-based non-adaptive beam alignment schemes have the potential to allow multiple \acp{UE} to perform initial access beam alignment in parallel whereas adaptive schemes are favourable in achievable beamforming gain.
This work introduces a novel deep learning based joint beam alignment scheme that aims to combine the benefits of adaptive, codebook-free beam alignment at the \ac{UE} side with the advantages of a codebook-sweep based scheme at the \ac{BS}.
The proposed end-to-end trainable scheme is compatible with current cellular standard signaling and can be readily integrated into the standard without requiring significant changes to it. Extensive simulations demonstrate superior performance of the proposed approach over purely codebook-based ones.
\end{abstract}

\acresetall

\section{Introduction}
The need for highly directional transmissions using beamforming techniques to overcome the large pathloss at mmWave frequencies requires systems operating at these frequencies to use large antenna arrays and highly directive \ac{BF}.
Due to the prohibitive cost and power consumption of fully digital systems operating at these high frequencies, mmWave systems often adopt analog or hybrid beamforming techniques.
These restrictions gives rise to the need for specific \ac{IA} strategies, often called \ac{BA} or \ac{BT}. 
There has been a great amount of work done in this area, many using beam-codebooks together with methods like compressed-sensing, machine learning, and Bayesian approaches \cite{Codebook_2014}\cite{ActiveLearning}\cite{AgileLink}.
Recently it has been shown that codebook-free, adaptive \ac{BA} might have performance benefits over non-adaptive and codebook based approaches \cite{BayesDNN} \cite{RNNE2E} \cite{DRLBA}.\\
Note that much of these works focus on the problem of one-sided \ac{BA}, where one side of the link sends omnidirectionally while the other side performs \ac{BA}.
While this can be extended in a straightforward way to two-sided \ac{BA} if one assumes a \ac{TDD} system, this approach might not be optimal in \ac{BA} latency or performance.
Other works, like in \cite{two_bandit_1} and \cite{two_bandit_2} perform joint, two-sided \ac{BA} by interpreting the \ac{BA} as a adversarial bandit problem using predefined beam codebooks. Recent work \cite{grid_free} introduces a joint \ac{DNN} based beam alignment scheme where first, both \ac{BS} and \ac{UE} sense the channel using a trainable set of (site-specific) beam-pairs, then the \ac{UE} feeds back the measurement results to the \ac{BS}, after which both sides determine their final, grid-free beam-pairs using \ac{DNN} and the measurement results obtained from the \ac{UE}.
Another approach \cite{ping_pong} extends the one-sided adaptive \ac{BA} scheme of \cite{ActiveLearning} to the joint, two sided \ac{BA} problem by introducing an adaptive, ping-pong pilot based approach where the role of \ac{TX} and \ac{RX} are switched after each sensing step, and demonstrates its superior performance over previous state-of-the-art methods in this domain. 
The work in this paper can be regarded as a combination of \cite{RNNE2E} (or, by extension, \cite{ping_pong}) and \cite{grid_free}, which aims to keep current cellular standard compatibility and benefits (like allowing multiple \ac{UE} to perform \ac{BA} simultaneously or cell discovery) of \cite{grid_free} while making use of the advantages provided by adaptive \ac{BA} in \cite{RNNE2E}.
The main contributions of this work are as follows:
\begin{itemize}
    \item We propose a new end-to-end unsupervised trained, \ac{DNN} based joint \ac{BA} algorithm which combines the benefits of a beam-sweep based \ac{BA} scheme at the \ac{BS} using learned beam codebooks and adaptive, codebook free \ac{BA} at the \ac{UE} side.
    \item The proposed method can be readily adopted into current cellular standards and can also be used in \ac{SA} architectures.
    \item We demonstrate the performance of our scheme in various simulations.
\end{itemize}
 \begin{figure}[!t]
   \vspace{-0.2cm}
  \centering
    \resizebox{1\columnwidth}{!}{\includegraphics{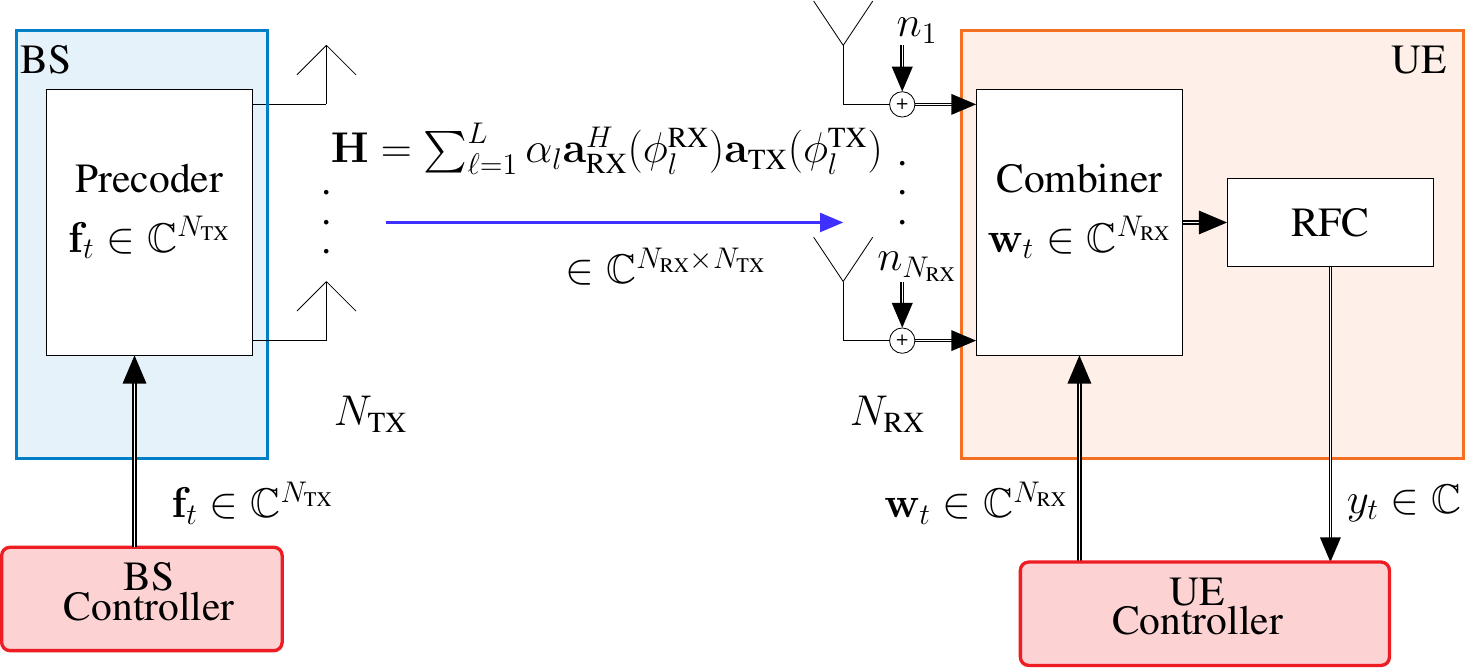}}
 \caption{\small System model of the proposed \ac{BA} algorithm. Note how at each timestep $t$, the beampatterns of \ac{BS} and \ac{UE} (i.e., $\mathbf{f}_t$ and $\mathbf{w}_t$) are determined by a respective \ac{CU}, i.e., joint \ac{BA} is performed.}
 \label{fig:system_model_ba}
 \vspace{-0.5cm}
\end{figure}
This paper is organized as follows: Section \ref{sec:prelim} describes the system model and introduces some important notations. 
Following this, Section \ref{sec:architecture} explains the proposed \ac{BA} method in more detail.
Section \ref{sec:results} investigates the performance of the proposed scheme through various experiments. 
Finally, Section  \ref{sec:conclusion} concludes the paper. 
\section{Preliminaries}
\label{sec:prelim}
In this paper, we consider the problem of joint, two-sided beam alignment in mmWave communications, i.e., the initial (downlink) communication between a \ac{BS} and an \ac{UE}, both equipped with an \ac{ULA} consisting of $N_{\text{RX}}$ and  $N_{\text{TX}}$ antenna elements respectively.
Note that in this setup, no initial knowledge about the channel between \ac{BS} and \ac{UE} is assumed.
In the rest of this paper, we assume that both the \ac{UE} and the \ac{BS} are equipped with only a single \ac{RF} chain, i.e., analog beamforming and that the transmission channel between UE and BS is static, i.e. constant during each run of the \ac{BA} algorithm.
Furthermore, it is assumed that both the \ac{BS} and the \ac{UE} are controlled by \acp{CU} which at each discrete timestep $t$, dictate the current precoding vector $\mathbf{f}_t \in \mathbb{C}
^{N_\text{TX}}$, and the combining vector $\mathbf{w}_t \in \mathbb{C}
^{N_\text{RX}}$ at the \ac{BS} and \ac{UE} respectively.
Also, the \ac{CU} at the \ac{UE} also receives the complex valued received symbol $y_t \in \mathbb{C}$ and the current \ac{BS} beam index $x_t \in \mathbb{Z}$ at each timestep as input. The system model is also depicted in Fig. \ref{fig:system_model_ba}.
\subsection{System Model}
As the mmWave channel can be regarded as being spatially sparse, we assume the widely used geometric channel model \cite{channel_model} for the mmWave channel description. 
Using this model, the channel can be described by:
\begin{equation}
    \mathbf{H} = \sum_{\ell = 1 }^{L} \alpha_l \mathbf{a}_{\text{RX}}^H(\phi^{\text{RX}}_ {\ell}) \mathbf{a}_{\text{TX}}(\phi^{\text{TX}}_{\ell}) \in \mathbb{C}^{N_{\text{RX}} \times N_{{\text{TX}}}}
    \label{eq:channel}
\end{equation}
for a channel with $L$ paths, complex path gain $\alpha_{\ell} \sim \mathcal{CN}(0,1)$, \ac{AoA} $\phi^{\text{RX}}_{\ell} \sim \mathcal{U}[-\frac{\pi}{2},\frac{\pi}{2}]$ and \ac{AoD} $\phi^{\text{TX}}_{\ell} \sim \mathcal{U}[-\frac{\pi}{2},\frac{\pi}{2}]$ of path $\ell$. $\mathbf{a}(\phi)$ describes the array response of the \ac{ULA} for angle $\phi$ and is given by
\begin{equation}
\mathbf{a}(\phi) = (1, e^{j \pi \cos(\phi)}, \dots, e^{j\pi(N-1)\cos(\phi)}) \in \mathbb{C}^{N}
\end{equation}
with $N$ antennas, and under the assumption of an element spacing of $\frac{\lambda}{2}$. Note that an \ac{ULA} was chosen as the antenna array geometry due to its simplicity, but other forms like \acp{UPA} are readily integratable as the proposed method is agnostic on the array geometry. Also, as it is assumed that the \ac{BS} uses only one \ac{RF} chain and sends a constant stream of pilot symbols with normalized power $P_{\text{TX}} = 1$, the received signal at the \ac{UE} at timestep $t$ can be described using
\begin{equation}
    y_t = \mathbf{w}_t^H \mathbf{H} \mathbf{f}_t+\mathbf{w}_t^H \mathbf{n}_t \, \in \mathbb{C}
    \label{eq:system_equation}
\end{equation}
with $y_t$ being the complex received symbol at the \ac{UE}, $\mathbf{w}_t$ being the $N_{\text{RX}}$ dimensional complex analog combining vector, $\mathbf{f}_t$ being the $N_{\text{TX}}$ dimensional complex analog precoding vector, and noise vector $\mathbf{n}_t \sim \mathcal{CN}(0,\sigma_n^2 \mathbf{I}_{N_{\text{RX}}})$, all for timestep $t$.
The receive \ac{SNR} after combining for a given $\mathbf{w}_t$ and $\mathbf{f}_t$ is determined as
\begin{equation}
   \ac{SNR}_{\text{RX}} = \frac{| \mathbf{w}_{t}^H \mathbf{H} \mathbf{f}_{t} |^2}{| \mathbf{w}_{t}^H \mathbf{n} |^2}
\end{equation}
and the per-antenna $\ac{SNR}_{\text{ANT}} = \ac{SNR}$ is defined as $\ac{SNR} = \frac{1}{\sigma_n^2}$. The (link) satisfaction probability is defined as
\begin{equation}
   P_{\text{sat}} = \mathbb{E}_{\mathbf{H}} [\mathbbm{1}_{\mathcal{H}_{\text{sat}}}(\mathbf{H})]
\end{equation}
where $\mathbbm{1}$ denotes the indicator function and 
\begin{equation}
    \mathcal{H}_{\text{sat}}= \{ \mathbf{H} : \ac{SNR}_{\text{RX}} > \text{SNR}_{\text{threshold}}\}
\end{equation}
is the set of all channel matrices $\mathbf{H}$ for which the receive \ac{SNR} at the last timestep is above a threshold value $\text{SNR}_{\text{threshold}}$.
In verbal terms, $P_{\text{sat}}$ measures the ratio of channels for which the receive \ac{SNR} of the considered \ac{BA} scheme satisfies a certain \ac{SNR} (throughput) threshold.
If not mentioned otherwise, in the following, ``beamforming gain'' refers to the expression $| \mathbf{w}_{t}^H \mathbf{H} \mathbf{f}_{t} |^2$.
As the channel matrix $\mathbf{H}$ between the \ac{UE} and \ac{BS} is not assumed to be known initially, the goal of the proposed scheme is to first collect as much implicit \ac{CSI} about $\mathbf{H}$ at the \ac{UE}, and then, use this implicit \ac{CSI} both at the \ac{UE} and \ac{BS} to construct the optimal grid-free final beampair.
\section{The proposed NN based solution}
\label{sec:architecture}
\begin{figure}[!t]
  \centering
   \vspace{0.15cm}
    \resizebox{1\columnwidth}{!}{\includegraphics{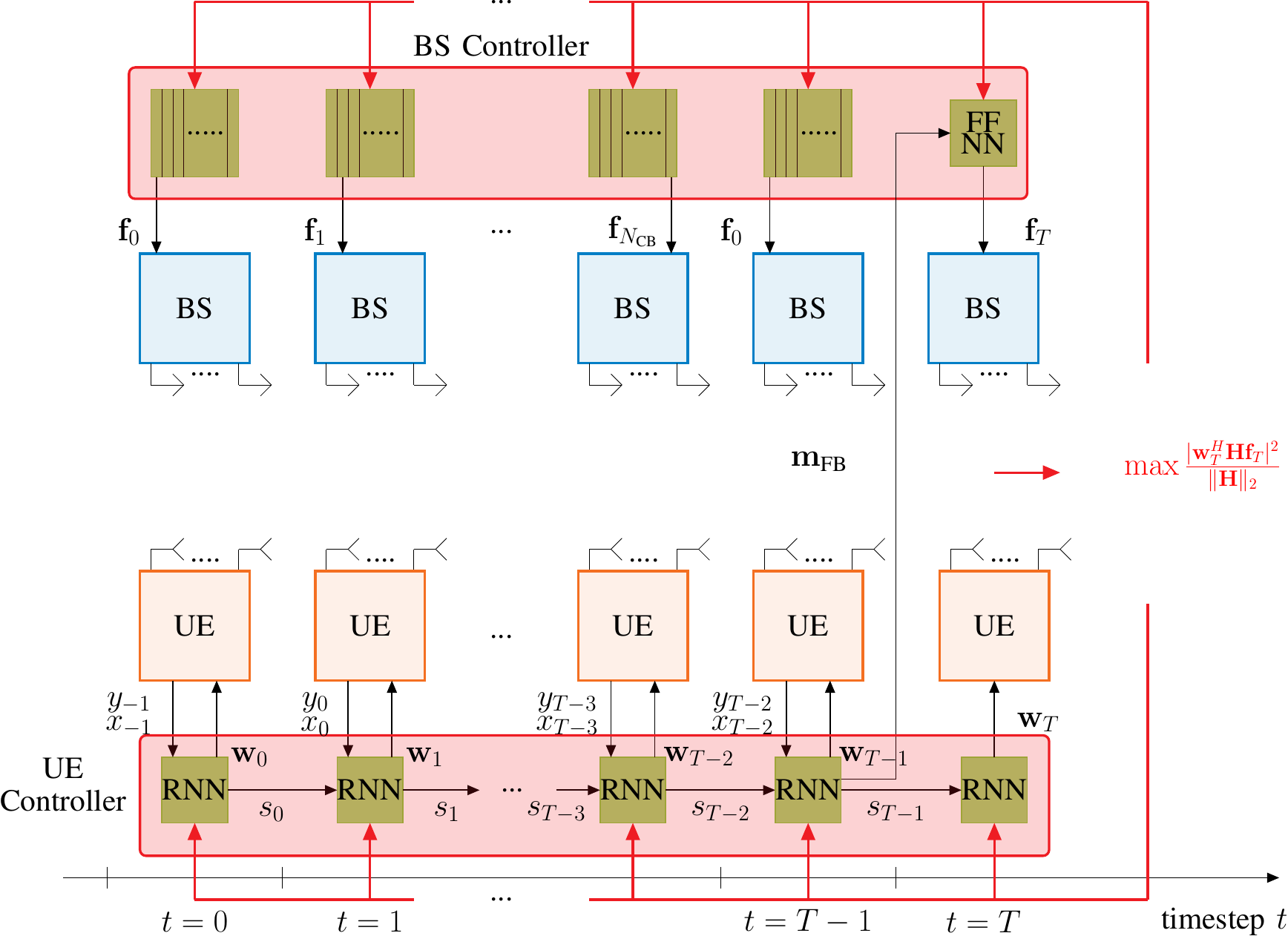}}
 \caption{\small Unrolled depiction of the proposed joint \ac{BA} algorithm. For $0 \leq t \leq T-1$ the \ac{BS} selects its beam based on its learnable codebook, but for $t = T$, the beam is determined codebook free, based on the received feedback $\mathbf{m}_{\text{FB}}$ from the \ac{UE}. Note that green shaded boxes represent trainable components. The red arrow indicates the flow of the gradient used for updating the trainable parameters in the scheme.}
 \label{fig:system_training}
 \vspace{-0.55cm}
\end{figure}
The objective of our proposed scheme is to use the advantages of a codebook-free adaptive \ac{BA} while keeping the practical benefits that a \ac{BS} using repeating beam-sweeps offers.
The protocol can be divided into three different stages:
\begin{enumerate}
    \item \textbf{Channel sensing stage:} The \ac{UE} initiates the \ac{BA} process and performs $T$ adaptive sensing steps. During this phase, the \ac{BS} exhaustively and repeatedly sweeps through its own codebook with $N_{\text{CB}}$ entries (note that in general, $T$ can be larger than $N_{\text{CB}}$). Like in current cellular standards, each \ac{BS}-beam can be transmitted using \acp{RS}, like \ac{SSB} blocks. Furthermore it is also assumed that the \ac{UE} is always able to extract the beam index out of every \ac{SSB} beam, irrespective of the \ac{SNR} after combining.
    \item \textbf{Feedback determination and final UE beamforming:} After the $T$ sensing steps, the \ac{UE} determines 
    \begin{enumerate}
        \item its own predicted best combining vector $\mathbf{w}_{T}$
        \item the feedback message $\mathbf{m}_{\text{FB}} \in \mathbb{R}^{N_{\text{FB}}}$ for the \ac{BS}
    \end{enumerate}
    The \ac{UE} then feeds back $\mathbf{m}_{\text{FB}}$ to the \ac{BS}. This can be either done over a lower frequency sidechannel in \ac{NSA} systems, or, using $\mathbf{w}_{T}$ for \ac{SA} systems (i.e., here the \ac{UE} transmits and the \ac{BS} receives).
    \item \textbf{Final BS beamforming:} The \ac{BS} uses $ \mathbf{m}_{\text{FB}}$ to determine its own predicted best combining vector $\mathbf{f}_{T}$ to use for subsequent communications with the \ac{UE}.
\end{enumerate}
The unrolled depiction of the proposed algorithm can also be seen in Fig. \ref{fig:system_training}. 
To utilize the various benefits that \ac{NN} can offer, we use three different kinds of learnable networks: The \ac{UE} sided \ac{RNN} (N1), the final beam-mapping network at the \ac{BS} (N2), and the learnable beam-codebook at the \ac{BS} (N3). The purpose of these \ac{NN} is as follows:
\begin{itemize}
    \item \textbf{N1 (\ac{RNN}):} Sensing network with the goal of obtaining as much \ac{CSI} information as possible about the channel between \ac{BS} and \ac{UE}, based on the measurement sequence of $\mathbf{w}_t$, $y_t$ and $x_t$. The obtained information is stored in the internal state $\mathbf{s}_t$. At each timestep $t$, it outputs a combining vector $\mathbf{w}_t$, and for $t=T-1$, it additionally determines a feedback massage $\mathbf{m}_{\text{FB}}$ for the \ac{BS}. Also, at the last timestep of the \ac{BA} process, it outputs the estimated best combining vector $\mathbf{w}_{T}$ for the \ac{UE}.
    \item \textbf{N2 (\ac{FNN}):} Network to map the feedback massage $\mathbf{m}_{\text{FB}}$ of N1 to the final beampattern $\mathbf{f}_{T}$ at the \ac{BS}.
    \item \textbf{N3 (parameter matrix):} Learnable beam-codebook at the \ac{BS} for possible enhancements over conventionally used codebooks and hardware impairments. This codebook is implemented as a $2 \cdot N_{\text{TX}} \times N_{\text{CB}}$ dimensional trainable parameter matrix as discussed in \cite{grid_free}.
\end{itemize}
N1 is contained inside the \ac{UE} \ac{CU}, whereas N2 and N3 are both located inside the \ac{BS} \ac{CU}.
Formally, the objective of the proposed method is to maximize the following objective function via stochastic gradient-ascent:
\begin{align}
   \max_{\boldsymbol{\theta}_1,\boldsymbol{\theta}_2,\boldsymbol{\theta}_3,\boldsymbol{\theta}_4} \frac{|\mathbf{w}_{T}^H \mathbf{H} \mathbf{f}_{T}|^2}{\lVert \mathbf{H} \rVert_2}
 \label{eq:target}
\end{align}
i.e., to maximize the beamforming gain at the final timestep. The objective is normalized by the channel norm (using the Frobenius norm, denoted as $\lVert.\rVert_2$) to avoid overemphasis on \acp{UE} with good channels \cite{grid_free}. The maximization is subject to the following constraints:
\begin{itemize}
    \item Unit-norm beamforming vectors:
    \begin{align}
     \lVert  \mathbf{w}_{t}  \rVert_2 = 1  \, \forall \, t \nonumber\\
   \lVert  \mathbf{f}_{t}  \rVert_2 = 1  \, \forall \, t
       \label{eq:constr_1}
    \end{align}
 
     \item Adaptive \ac{UE} sensing vectors:
    \begin{align}
 \mathbf{w}_{t} = f_{\boldsymbol{\theta}_1}^{\text{UE}} (y_{< t}, x_{< t}, \mathbf{w}_{<t})  \, \forall \, t
    \end{align}
     \item Codebook based \ac{BS} precoding vectors in the first $T$ timesteps:
    \begin{align}
 \mathbf{f}_{t} = \mathbf{F}_{\boldsymbol{\theta}_4,[t+i] \text{mod} N_{\text{CB}},:}^{\text{BS}}  \, \forall \, 0 \leq t < T 
    \end{align}
     \item Codebook-free \ac{BS} precoding vector in the last timestep:
    \begin{align}
  \mathbf{f}_{T} =   f_{\boldsymbol{\theta}_3}^{\text{BS}}(\mathbf{m}_{\text{FB}})
    \end{align}
      \item Real valued feedback message $\mathbf{m_{\text{FB}}}$:
    \begin{align}
 \mathbf{m}_{\text{FB}} =  f_{\boldsymbol{\theta}_2}^{\text{UE}}(y_{< T-1}, x_{< T-1}, \mathbf{w}_{< T-1})
 \label{eq:feedback}
    \end{align}
\end{itemize}
where the superscript ${\text{BS}}$ and ${\text{UE}}$ denotes that the component is used in the transmitter (BS) and receiver (UE), respectively, $y_{-1} = 0$, $x_{-1} = -1$, $\mathbf{w}_{-1} = \mathbf{0}$. $i \in [0,N_{\text{CB}}]$ denotes a random starting index in the \ac{BS} sided beam codebook to potentially enable the \ac{UE} to learn a \ac{BA} algorithm without having to assume a fixed codebook index at the \ac{BS}, which could reduce the delay of the \ac{IA} phase in application scenarios.
The nonlinear, parameterized functions $f_{\boldsymbol{\theta}_1}^{\text{UE}}: \mathbb{C}^{t} \times \mathbb{R}^{t} \times \mathbb{C}^{N_{\text{RX}}t} \rightarrow \mathbb{C}^{N_{\text{RX}}}$, $f_{\boldsymbol{\theta}_2}^{\text{UE}}: \mathbb{C}^{T-1} \times \mathbb{R}^{T-1} \times \mathbb{C}^{N_{\text{RX}}(T-1)} \rightarrow \mathbb{R}^{N_{\text{FB}}}$ are implemented using N1, $f_{\boldsymbol{\theta}_3}^{\text{BS}}: \mathbb{R}^{N_{\text{FB}}} \rightarrow \mathbb{C}^{N_{\text{TX}}}$ by N2 and $ \mathbf{F}_{\boldsymbol{\theta}_4,[t+i] \text{mod} N_{\text{CB}},:}^{\text{BS}} \in \mathbb{C}^{N_{\text{CB}} \times N_{\text{TX}}}$ is the parameterized codebook N3. $\boldsymbol{\theta}_1,\boldsymbol{\theta}_2,\boldsymbol{\theta}_3,\boldsymbol{\theta}_4$ are the vectorized representations of all the learnable parameters in the respective \ac{NN}. Furthermore, we introduce three different configurations of our proposed \ac{BA} scheme, using an increasing amount of trainable components:
\begin{itemize}
    \item \textbf{C1:} Here only N1 is assumed to be used, i.e., only the \ac{UE} side is modified and the BS uses a codebook determined via a conventional optimization method \cite{codebook}. Consequently, in this case the feedback from \ac{UE} to \ac{BS}, namely $\mathbf{m}_{\text{FB}}$, is a single number representing the estimated best \ac{BS} side beam index and Eq. \ref{eq:feedback} is slightly modified to:
    \begin{equation}
        m_{\text{FB}} = \max_k [\boldsymbol{\sigma}(f_{\boldsymbol{\theta}_2}^{\text{UE}}(y_{< T-1}, x_{< T-1}, \mathbf{w}_{< T-1}))_k]
    \end{equation}
    where $\boldsymbol{\sigma}: \mathbb{R}^{N_{\text{CB}}} \rightarrow (0,1)^{N_\text{CB}}$ is the $N_{\text{CB}}$ dimensional softmax function mapping the $N_{\text{CB}}$ dimensional feedback output of N1 to a probability vector. In order to train the learnable parameters of $f_{\boldsymbol{\theta}_2}^{\text{UE}}$, a cross-entropy loss is applied with the one-hot encoding of the optimal \ac{BS} beam index as label (ground truth) at timestep $t = T-1$.
    \item \textbf{C2:} This variant is equivalent to variant C1 with the addition of using the learnable map N2.
    \item \textbf{C3:} Here, everything is the same as C2 with the addition of using the learned beam codebook N3, i.e., here all components are learnable.
\end{itemize}
Note that variant C1 introduces only a \ac{UE}-sided modification to the \ac{BA} process and the \ac{BS} performs the standard conventional codebook based beam sweeps without any modifications, as also done in current cellular standards. The three introduced configurations are also summarized in Table \ref{tab:configurations}.
\begin{table}[t!]
\centering
 \vspace{0.1cm}
\begin{tabular}{| m{5.5em} | m{7em}| m{6.5em} | m{6.5em} |} 
 \hline
 \textbf{Configuration} &  \textbf{\ac{BS} beam codebook} &  \textbf{\ac{BS} final beam} & \textbf{\ac{UE} beampatterns} \\ [0.5ex] 
 \hline\hline
C1 & conventional & conventional & learned\\ 
 \hline
C2 & conventional & learned & learned\\
 \hline
C3 & learned & learned & learned\\
 \hline
\end{tabular}
 \caption{Summary of the configurations for the proposed scheme.} \label{tab:configurations}
 	\vspace{-0.5cm}
\end{table}
As previously mentioned, the proposed method can be trained straightforward in an end-to-end fashion in the following way:
First, sample a random channel matrix according to Eq. \ref{eq:channel}. Then, unroll the \ac{BA} algorithm over time using this channel matrix and using equations \ref{eq:system_equation}, \ref{eq:constr_1} - \ref{eq:feedback}.
Finally, update the trainable weights involved in the computation, $\boldsymbol{\theta}_1,\boldsymbol{\theta}_2,\boldsymbol{\theta}_3,\boldsymbol{\theta}_4$, using \ac{BPTT} and gradient ascent to maximize the objective of Eq. \ref{eq:target}.
To stabilize training, the update is averaged over a batch of $N_{\text{batch}}$ entries.
Due to the modularity of some trainable components (e.g. the learnable codebook $N3$) of the proposed scheme, one can in theory also imagine to train some components independently from each other, e.g. through \ac{DRL}.
\section{Numerical Results}
\label{sec:results}
\begin{figure}[!t]
\centering
\resizebox{1\columnwidth}{!}{\includegraphics{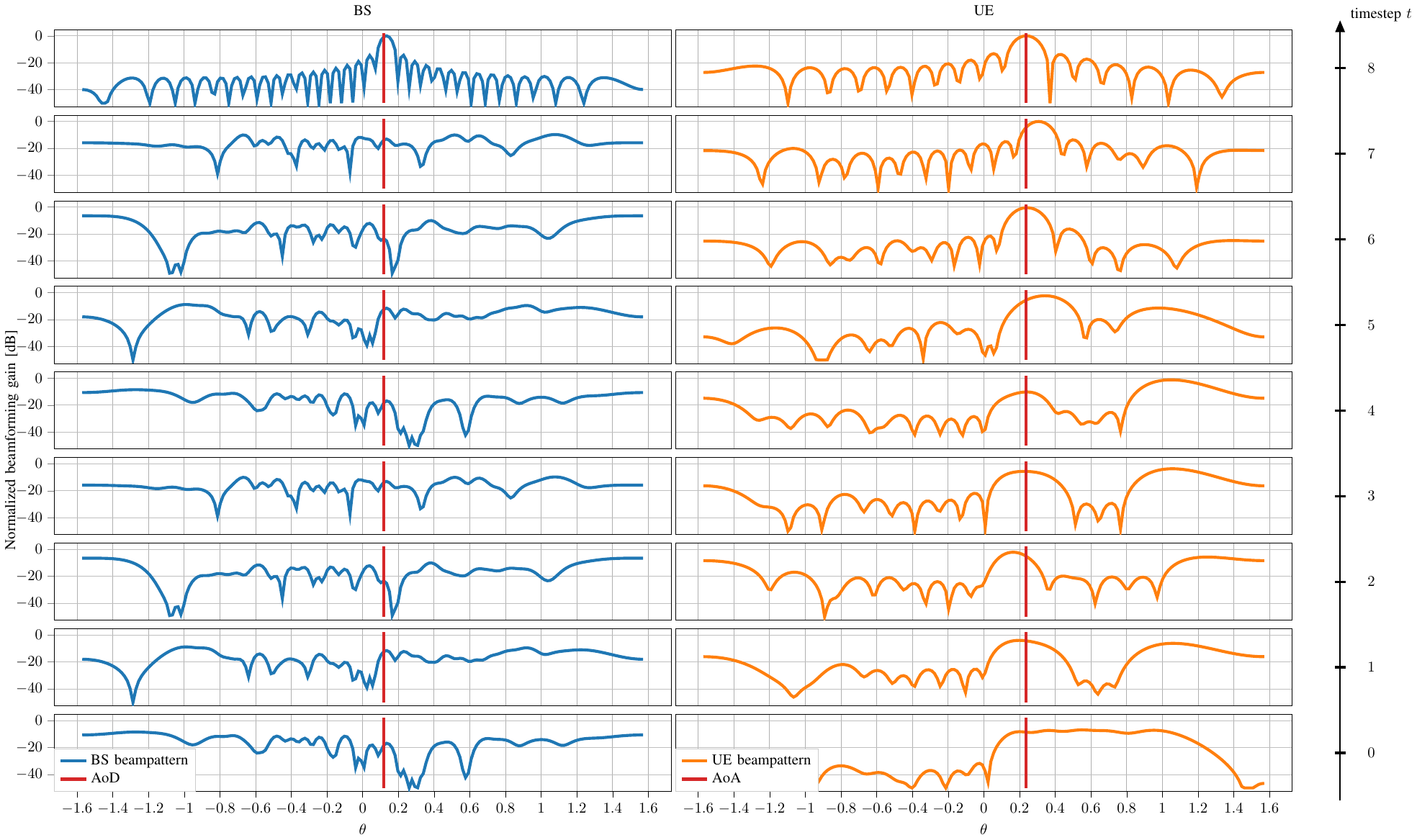}}
\caption{Unrolled visualization of the learned beampatterns for the proposed joint \ac{BA} algorithm for a specific channel realization.}
\label{fig:ba_process}
\vspace{-0.6cm}
\end{figure}
Our proposed method will be compared to the following baselines:
\begin{itemize}
    \item \textbf{Exhaustive Search:} Both the \ac{BS} and \ac{UE} exhaustively sweep through a equal beam-width-codebook generated by a conventional optimization algorithm \cite{codebook} after which the beam-pair obtaining the highest power is chosen. $M_{\text{TX}}$ ($M_{\text{RX}}$) denotes the number of beampatterns in the BS (UE)-sided codebook. For fairness of comparison, $M_{\text{TX}} \cdot M_{\text{TX}} = T$.
    \item \textbf{Maximum ratio transmission (MRT) and \ac{MRC}:} This is the theoretical upper bound for the given scenario where the \ac{BS} uses the MRT precoding vector and the \ac{UE} the \ac{MRC} combining vector.
    \item \textbf{Joint, non-adaptive ($\text{DNN}_{\text{NOA}}$):} A \ac{DNN}-based scheme as introduced in \cite{grid_free}, where both the \ac{BS} and \ac{UE} jointly sweep through a set of learned beampairs, after which the \ac{UE} feeds back the measurement results and both sides determine a grid-free final beampair based on these measurement results.
     \item \textbf{Joint, adaptive ($\text{RNN}_{\text{A}}$):} A \ac{RNN}-based ping-pong scheme as introduced in \cite{ping_pong}, where both the \ac{BS} and \ac{UE} jointly perform adaptive, codebook-free \ac{BA}, and switch the role of \ac{TX} and \ac{RX} after each timestep. Note that here one timestep $t$ refers to one ``channel use'', i.e., in each timestep one side transmits and the other side receives.
\end{itemize}
Note that $\text{DNN}_{\text{NOA}}$ can be seen as an approach where both parties of the \ac{BA} procedure are codebook-bound (except for the last beam-pair), non-adaptive and in $\text{RNN}_{\text{A}}$, both sides perform codebook-free, adaptive \ac{BA}, whereas our approach has one side (i.e., the \ac{BS}) non-adaptive, codebook-based and the other side (i.e., the \ac{UE}) codebook-free, non-adaptive.
As the focus here lies on comparing the performance of the high-level schemes, the loss functions of both $\text{DNN}_{\text{NOA}}$ and $\text{RNN}_{\text{A}}$ were set to the same loss-function also used for the proposed method (without any auxiliary losses), i.e. the implementations used in this paper slightly differ from those of the original proposals.
Also note that for ease of explanation and simplicity, both the \ac{UE} of the proposed method and $\text{RNN}_{\text{A}}$ do not use a separate \ac{DNN} to determine the final beam(s) at the last timestep.
An overview over the high level characteristics of the baselines is also shown in Table \ref{tab:comparison}.
In the following experiments, if not mentioned otherwise, $L=3$, $N_{\text{TX}} = 32$, $N_{\text{RX}} = 16$,  $N_{\text{CB}} = 8$, batch size $N_{\text{batch}} = 1024$, and the training \ac{SNR} is set to $10$ dB.
Also, the threshold value $\text{SNR}_{\text{threshold}}$ is set to $20$ dB and the dimension of $\mathbf{m}_{\text{BF}}$, $N_{\text{FB}}$, is set to $16$.
Furthermore, the amount of trainable parameters in the tested \ac{NN} based schemes is set to be roughly $500 K$. The \acp{RNN} used for the UE side (N1) in this scheme consists of two layers of \acp{GRU} \cite{GRU} whereas components (e.g. N2) employing feedforward networks consist of two to three layers of fully connected \ac{DNN}.

Fig \ref{fig:ba_process} aims to provide a visualization of the learned \ac{BA} algorithm for $L=1$, $T=8$ and $N_{\text{CB}}=4$ (i.e., $8$ \ac{UE}-sensing steps), and for configuration C3.
Plotted are the normalized beamforming gains of the beampatterns (calculated using $ | \mathbf{v}^H\mathbf{a}(\phi) |^2$ with $\phi \in [-\frac{\pi}{2},\frac{\pi}{2}]$ and combining (precoding) vector $\mathbf{v}$) used by the \ac{BS} (marked in blue) and \ac{UE} (marked in orange) during the \ac{BA} process.
It can be clearly seen that in the first stages of the \ac{BA} process, the \ac{UE} senses the channel with relatively wide beampatterns, whereas for the later timesteps of the algorithm, the search area of the \ac{UE} is narrowed down and more energy is focused in certain directions.
Also note the repeating \ac{BS}-sided beampatterns in the first $8$ timesteps, caused by its codebook-based beamsweep.

\begin{figure}[!t]
	\centering
	\vspace{-0.1cm}
\begin{subfigure}[b]{0.5\textwidth}
\includegraphics{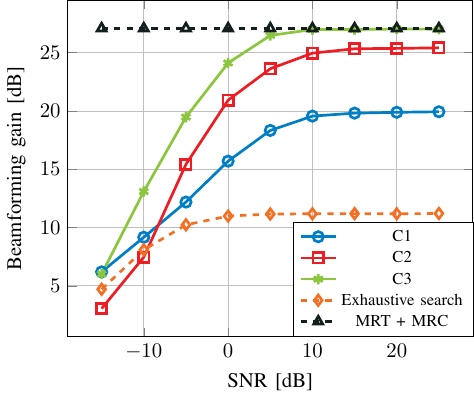}
   \label{fig:1} 
\end{subfigure}
\begin{subfigure}[b]{0.5\textwidth}
\includegraphics{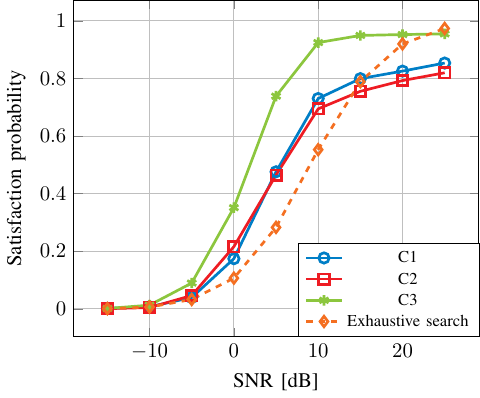}
   \label{fig:2}
 \end{subfigure}
\caption{\small Comparison of the performance impact of the various \acp{NN} used in the proposed scheme, top: beamforming gain versus \ac{SNR}, bottom: satisfaction probability versus \ac{SNR}.}
	\vspace{-0.62cm}
	\label{fig:comparison_components}
\end{figure}

In a first experiment, we want to investigate the effects of the various learnable components on the performance of the proposed \ac{BA} procedure, i.e., we want to investigate the performance of C1, C2 and C3, as described in the previous section.
The results are shown in Fig. \ref{fig:comparison_components}, $T=16$.
One can observe that all three tested variants outperform the exhaustive search by at least $10$ dB and that introducing an increasing amount of learnable components into the scheme improves the performance.
Also remarkable is the large benefit in regards to satisfaction probability when introducing a learnable beam-codebook at the \ac{BS}.

\begin{figure}[!t]
	\centering
	
\resizebox{1.1\columnwidth}{!}{\includegraphics{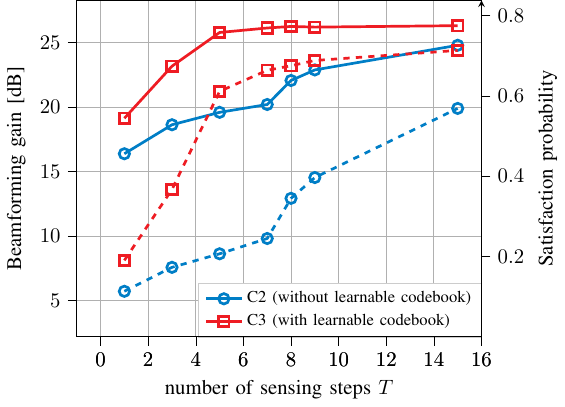}}
\caption{\small Comparison of the performance impact of varying $T$ for fixed $N_{\text{CB}} = 8$ for the proposed scheme. Solid lines: beamforming gain versus $T$, dashed lines: satisfaction probability versus $T$.}
	\label{fig:comparison_codebook}
\end{figure}

Another aspect we want to examine is the impact of $T$, i.e., the number of sensing steps, on the performance for a fixed $N_{\text{CB}}$, i.e., the potential benefit of additional beam alignment steps after the \ac{UE} has observed all $N_{\text{CB}}$ beams of the \ac{BS}.
To also investigate the influence of the learnable codebook on the performance (especially if $T \leq N_{\text{CB}}$), the experiment is performed for both the variants C2 and C3.
For this experiment, the parameters of the system are chosen to be identical to those of the previous experiment, and the per-antenna \ac{SNR}, $\ac{SNR}_{\text{ANT}}$, is held constant to $5$ dB. The results are displayed in Fig. \ref{fig:comparison_codebook}.
One can observe that while the performance of both tested variants increases steadily by increasing $T$, the gap in their performance between allowing one and more sensing steps also increases.
Also visible is the general large increase both in the achievable beamforming gain and satisfaction probability if a learned beam-codebook is introduced into the scheme, especially for small $T<N_{\text{CB}}$.

\begin{figure}[!t]
	\centering
	\vspace{-0.5cm}
\resizebox{1.1\columnwidth}{!}{\includegraphics{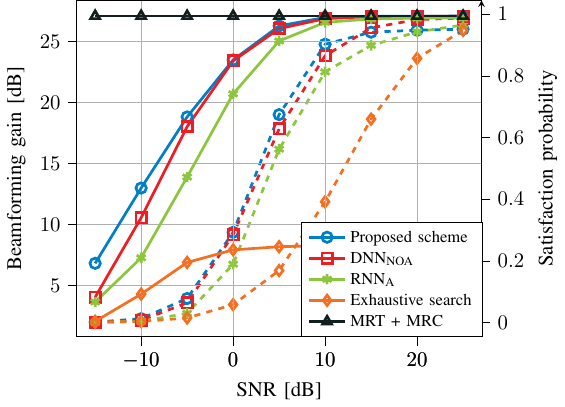}}
\caption{\small Performance of different \ac{BA} schemes for fixed $T = 8$ and different values of testing $\ac{SNR}_{\text{ANT}}$. Shown are the beamforming gain (solid) and the satisfaction probability (dashed).}
	\vspace{-0.55cm}
	\label{fig:comparison_baselines}
\end{figure}

\begin{figure}[!t]
	\centering
	\vspace{-0.15cm}
\resizebox{1\columnwidth}{!}{\includegraphics{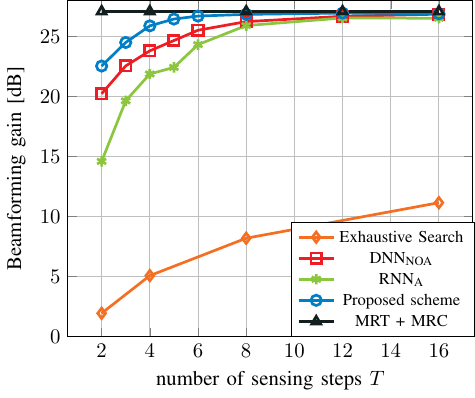}}
\caption{\small Beamforming gain versus number of sensing steps $T$ of different \ac{BA} schemes for fixed $\ac{SNR}_{\text{ANT}} = 5$ dB.}
	\label{fig:gains_vs_t}
		\vspace{-0.1cm}
\end{figure}

In two further experiments, the performance of the proposed scheme in comparison to several relevant baselines is investigated.
To keep the comparison fair, in this experiment, the starting beam-codebook index $i$ of the proposed scheme is fixed to $0$.
In the first experiment, the amount of sensing steps, $T$, is fixed to $8$ for all schemes. The results are shown in Fig. \ref{fig:comparison_baselines}.
As can be seen, all tested \ac{DNN}-based schemes outperform the exhaustive search baseline in average beamforming gain by orders of magnitude.
Our proposed scheme slightly beats all tested baselines both in achieved beamforming gain and satisfaction probability in this experiment, even $\text{RNN}_{\text{A}}$.
The performance improvements of our proposed scheme in comparison to $\text{DNN}_{\text{NOA}}$ might be attributed to the adaptiveness at the \ac{UE} side whereas the difference to $\text{RNN}_{\text{A}}$ might be caused by the differing feedback schemes.
These observations also hold in a next experiment: Here we now fix the training and testing \ac{SNR} to $\ac{SNR}_{\text{ANT}} = 5$ dB, and vary the amount of sensing steps $T$.
For the learnable schemes, a different \ac{NN} is trained for each $T$, and for the proposed scheme, the codebook-size at the \ac{BS}, $N_{\text{CB}}$, is set to $8$. 
Fig. \ref{fig:gains_vs_t} shows the results.
Here again, the differences in the achieved average beamforming gain between the tested schemes become visible, most likely due to the previously mentioned reasons.
Also noteworthy is that all tested learned schemes at $T=2$ already outperform the exhaustive search baseline at $T=16$ sensing steps in regard to achieved average beamforming gain.
Note however that the exhaustive search-based scheme has the lowest computational complexity out of the tested schemes and further investigations into the performance-complexity trade-off of the \ac{NN} based schemes might be beneficial. 
Furthermore, due to the codebook-sweep based \ac{BS} in our proposed scheme, in theory many \ac{UE} can perform \ac{BA} in parallel. 
The overhead of the various methods for alignment of $K$ \ac{UE} in terms of required pilots and feedback signals is also shown in Tab. \ref{tab:comparison}.
\begin{table}[t!]
 \vspace{0.1cm}
\centering
\begin{tabular}{| m{4em} | m{5.5em}| m{5.5em} | m{4em} | m{5.5em} |} 
 \hline
 \textbf{\ac{BA} scheme} & \textbf{\ac{BS} \ac{BA} method} &  \textbf{\ac{UE} \ac{BA} method} & \textbf{Beam sweeping overhead} &  \textbf{Feedback overhead} \\ [0.5ex] 
 \hline\hline
  \textbf{Exhaustive search} & NOA, SEQ & NOA,SEQ & $M_{\text{TX}} M_{\text{RX}}$ & $K$ beam indices\\ 
 \hline
  \textbf{$\text{DNN}_{\text{NOA}}$} & NOA, JOINT & NOA, JOINT & $T$ & $K \cdot T$ power values  \\
 \hline
  \textbf{$\text{RNN}_{\text{A}}$} &  A, JOINT & A, JOINT & $K \cdot T$ & $0$  \\
 \hline
  \textbf{Proposed} &  NOA, JOINT & A, JOINT & $T$ & $K N_{\text{FB}}$ real valued numbers \\ [1ex] 
 \hline
\end{tabular}
 \caption{Overview over the compared approaches together with their beam sweeping and feedback overhead for $K$ \ac{UE}. ``NOA'' = non-adaptive,``A'' = adaptive, ``SEQ'' = sequential, ``JOINT'' = joint.} \label{tab:comparison}
 	\vspace{-0.5cm}
\end{table}
Note that experiments showed that a small number of $N_{\text{FB}}$ seems to be sufficient to enable good performance and thus, $N_{\text{FB}}$ could be expected to stay well below $T$.
\section{Conclusion}
We propose a novel deep learning based initial access joint beam alignment scheme with combines the advantages offered by adaptive, codebook free beam alignment at the \ac{UE} and codebook based beam sweeping at the \ac{BS}.
The introduced scheme is trainable in an end-to-end fashion and is shown to offer superior performance to purely codebook-sweep based methods. 
It is designed in such a way that it can be readily applied in existing beam sweeping-based frameworks without requiring modifications to the alignment process. 
Furthermore, our scheme does not rely on starting on a fixed codebook beam from the BS, and thus, can start the beam alignment process anytime without having to wait for a specific BS codebook beam to appear. 
There are various possible future directions this line of research could be extended: For example, one could try to train the \ac{UE} using the deep reinforcement learning framework to adaptively terminate the \ac{BA} process depending on the channel and the target receive \ac{SNR} without having a fixed $T$ in order to save redundant sensing steps.
One could also investigate and study the behaviour and possible extension of the proposed scheme to situations where the \ac{BS} employs codebooks not encountered during training.
\label{sec:conclusion}

\bibliographystyle{IEEEtran}
\bibliography{IEEEabrv,references}

\end{document}